\documentstyle[aps,prl,multicol,epsf]{revtex}
\renewcommand{\v}[1]{{\bf #1}}

\newcommand{\s}{{\sigma}}

\newcommand{\rb}{{\bar{\rho}}}

\newcommand{\w}{{\omega}}
\newcommand{\zh}{{\hat{z}}}

\newcommand{\gr}{{\nabla}}
\def\eqa{\begin{eqnarray}}
\def\eea{\end{eqnarray}}

\newcommand{\eq}{\begin{equation}}
\newcommand{\ee}{\end{equation}}
\newcommand{\nn}{\nonumber\\}
\newcommand{\Eq}[1]{Eq.~(\ref{#1})}

\newcommand{\p}{\partial}

\newcommand{\ra}{\rightarrow}

\newcommand{\al}{\alpha}

\newcommand{\rxx}{\rho_{xx}}
\newcommand{\rxy}{\rho_{xy}}

\newcommand{\sxy}{\sigma_{xy}}

\def\journal #1, #2, #3, 1#4#5#6{{\sl #1~}{\bf #2}, #3 (1#4#5#6) }

\begin{document}
\draft
\widetext

\title{An Unsettled Issue in the Theory of the Half-Filled Landau Level}
\author{Dung-Hai Lee}
\address{
Department of Physics, University of California at
Berkeley, Berkeley, CA 94720, USA \\ }

\maketitle

\widetext
\begin{abstract}

\rightskip 54.8pt

The purpose of this paper is to identify an unsettled
issue in the theory of the half-filled Landau level, and
state our point of view.
\end{abstract}


\begin{multicols}{2}


\narrowtext

Whether the half-filled Landau level can be described
in terms of a ``Fermi liquid''\cite{hlr} of composite
fermions has been controversial.\cite{hlr,lkgk,shankar,
lee,haldane,read,stern,stern1} (Hereafter we shall use the
phrase ``Fermi liquid'' in a loose sense. It does not
imply, e.g., a finite quasiparticle weight, but simply means
that the long wavelength/low energy current-current
correlation functions resemble those of a liquid of fermions
in {\it zero magnetic field}.)

This controversy is somewhat side-tracked by the
recent development of an alternative description.
\cite{shankar,lee,haldane,read} In this new description
$\nu=1/2$ is pictured as a liquid of
fermionic dipoles with each dipole being made up of a
composite fermion and a correlation hole. (This way of describing $\nu=1/2$
originates from an insightful paper by Read.\cite{read0})
Recent evidences suggest that this description has
similar infrared difficulty as the fermion Chern-Simons
theory.\cite{read,stern,stern1}

The following are the main points of this paper:

\begin{enumerate}
\item The fermion Chern-Simons theory is {\it not} equivalent
to the composite Fermi liquid theory of Halperin, Lee
and Read.\cite{hlr} The former is a general formulation, the
latter is a bold statement about the dynamics.
\item It can be shown that the neutral fermion (dipole) action in
Ref.\cite{shankar,lee,haldane,read} is the same as that for the
fermion Chern-Simons theory in the lowest Landau
level.\cite{shankar,haldane}
\item The heart issue is whether, within the fermion
Chern-Simons
theory, one can describe the
half-filled Landau level as composite fermions moving
in {\it zero} magnetic
field.
\item If the electron Hall conductivity is $e^2/2h$ and its
longitudinal resistivity is non-zero, then
a) the composite fermion Hall conductivity
($\s_{xy}^{CF}$) is $-\frac{e^2}{2h}$, and b) the
neutral fermion Hall conductivity
($\s_{xy}^N$) is zero. There is
strong evidence that real systems (which all
theories intend to describe)
do show $\s_{xy}=e^2/2h$ and
$\rxx>0$.\cite{hw} This is consistent with the
presence of
particle-hole symmetry.
\item $\s_{xy}^{CF}=-\frac{e^2}{2h}$ implies that
the (polarization) charge current carried by
neutral fermions does not have off-diagonal correlation
at $\v q=0$.
\item Despite their large (negative) Hall conductance,
the composite fermions do have some aspect of a
Fermi liquid - their transverse current-current
correlation resembles that of electrons in zero magnetic field.
This mixed behavior is due to the fact that the
composite fermion motion is the superposition of two different
types of dynamics: the guiding-center-like intra-dipole
dynamics, and the zero-field-like inter-dipole dynamics.
\end{enumerate}

\noindent{\bf{I. The composite Fermi liquid theory}}
\\
\rm

An amazing fact about $\nu=1/2$ is that aside from
a large Hall conductance the behavior of electrons near
$\nu=1/2$ resembles that near zero field. This
statement applies to magneto-transport data\cite{jiang}, as well as
other Fermi-surface resonance experiments.\cite{focus,saw}

Shortly after Willett {\it et al}'s discovery of an
anomaly in the acoustic wave propagation\cite{willet}, a {\it
very} novel idea was put forward by Halperin, Lee and
Read (HLR).\cite{hlr} For the reasons described below, we
shall refer to the HLR work as the
``composite Fermi liquid theory'' (CFLT). The CFLT rests
 on the fermion
Chern-Simons description. In this description one views
each electron as a composite fermion carrying two quanta
of {\it fictitious} magnetic flux (see Fig.1).\cite{jain}
Unlike the electrons, the composite fermions see {\it
two} different magnetic fields: the applied field $B$,
and the fictitious field $b$. While $B$ is space-time
independent, $b$ is solenoid-like and time dependent.

\begin{figure}[h]
\epsfysize=2.5cm\centerline{\epsfbox{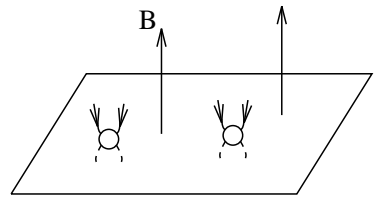}}
\vspace{20pt} \caption{Each electron is represented as a
composite fermion carrying two quanta of fictitious
magnetic flux. For later convenience the direction
of the
fictitious flux is chosen to be opposite to
$\hat{B}$.} \label{compferm}
\end{figure}

The virtue of the fermion Chern-Simons description is
that it suggests a novel mean-field theory. In this
mean-field theory one lets the {\it averaged} $b$
($\bar{b}=2\phi_0\rb$) cancel $B$. (Here $\rb$ is the average
electron/composite fermion density.) After the
cancellation the composite fermions see zero magnetic field hence
form a Fermi liquid. This mean-field theory is the basis of
Ref.\cite{hlr}.

In reality $b$ is space-time dependent, hence
can not cancel $B$ exactly. Attempts to go beyond mean-field theory
have not lead to a conclusive result. On this account
 HLR made a bold conjecture. They assert that the
cancellation between $b$ and $B$ is {\it not} spoiled
by the fluctuations beyond mean-field theory. Moreover they
assert that the sole effect of the fluctuations is to
renormalize the Fermi liquid parameters of composite
fermions.

An consequence of HLR's assertion is that the composite
fermion Hall conductance vanish: \eq
\s^{CF}_{xy}(\w=0,\v q=0)=0. \label{scf} \ee \Eq{scf}
lies at the heart of the issue we
shall discuss.

At this
point it is useful to contrast the mean-field theory
for $\nu=1/2$ with that for incompressible filling
factors.\cite{jain,lopez} The difference lies in the
fact that for incompressible filling factors the mean-field
theory predicts integer quantum Hall states, while for
$\nu=1/2$ it predicts a Fermi liquid. Since the former is
incompressible (hence does not have low energy $b$ fluctuations), the
statement that $b$ cancels part of $B$ is
asymptotically exact. The same can not be said about $\nu=1/2$,
 because the mean-field composite fermion state is
compressible.
\\

\noindent{\bf{II. The composite fermion Hall conductance}}
\\
\rm

Now let's come to the main issue - the validity of
\Eq{scf}. First let's recall the following {\it exact}
relation
between the electron and composite fermion resistivity
tensors ($\rho_{\al\beta}$  and $\rho^{CF}_{\al\beta}$):
\eq \rho_{\alpha\beta}=\rho_{\alpha\beta}^{CF}
+\epsilon_{\alpha\beta}\frac{2h} {e^2}. \label{8989} \ee In the
above $\rho_{\alpha\beta}^{CF}$ is defined so that $\s^{CF}_{\al\beta}\equiv
(\rho_{\alpha\beta}^{CF})^{-1}_{\al\beta}$ is the
conductivity deduced from the {\it
statistical-gauge-propagator-irreducible} current-current
correlation function of composite fermions.
\cite{hlr,klz} As usual, in the presence of long-range
interaction, the irreducible current-current correlation
describes the particle response to the total (i.e. external+internal) field.

The physics of \Eq{8989} is the fact that the
 Hall voltage seen by the composite fermions differs
 from that seen
by the electrons by an amount equals to $2\frac{h}{e^2}\times I$.
This difference comes from the fact that in the
composite fermion representation (Fig.2) there is a flux
current $I_{\phi} =2\frac{hc}{e}\frac{I}{e}$ in addition
to the charge current $I$. This flux current generates an
extra transverse voltage equals to
$\frac{1}{c}I_{\phi}=2\frac{h}{e^2}I$.

\begin{figure}[h]
\epsfysize=2cm\centerline{\epsfbox{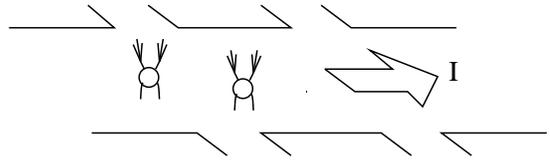}}
\vspace{20pt} \caption{Hall conduction from the composite
fermion point of view}
\label{hallgeom}
\end{figure}

As a result the longitudinal ($V_L, V^{CF}_L$) and
Hall ($V_H, V^{CF}_H$) voltages seen by the electron and the
composite fermion are related by

\eqa &&V_L=V_L^{CF}\nn &&V_H=V_H^{CF}+2\frac{h}{e^2}I.
\label{vol} \eea After dividing both sides of \Eq{vol} by
$I$ one obtains \Eq{8989}.

Next, we discuss another argument that is important for
setting up the issue concerning \Eq{scf} - the particle-hole
symmetry.\cite{lkgk} In the absence of disorder, particle-hole
symmetry emerges at $\nu=1/2$ after the {\it
projection onto the lowest Landau level}. The presence of
such symmetry implies that \eq
\s_{xy}=\frac{e^2}{2h}.\label{sxy} \ee

A caricature of the proof\cite{lkgk} goes as follows.
Upon the particle-hole conjugation the electron conductivity
tensor transforms as \eqa &&\s_{xx}(\nu)=\s_{xx}^h(1-\nu)\nonumber
\\ &&\s_{xy}(\nu)=\frac{e^2}{h}-\s_{xy}^h(1-\nu). \label{ph} \eea
In the above $\s_{\alpha\beta}^h$ is the conductivity
tensor of {\it holes}. The physical meaning of Eq.(\ref{ph})
is clear - after particle-hole conjugation the new
vacuum is a full Landau level and the total current is the sum
of the Hall current carried by the full Landau level
and the current carried by the holes. At $\nu=1/2$ we have
$\nu=1-\nu=1/2$ and particle-hole symmetry. As the
result $\s_{\alpha\beta}^h(1-\nu)=\s_{\alpha\beta}(\nu)$, and
hence \Eq{sxy} holds.

In the presence of disorder particle-hole symmetry
can at most hold on average. If the probability distribution of
the disorder potential satisfies $P[V(\v x)]=P[-V(\v x)]$
we say that the disorder is particle-hole symmetric. It
is important to note that while \Eq{8989} holds for
general disorder, \Eq{sxy} is only true when the disorder is
particle-hole symmetric. In either case when there is
disorder we need to interpret $\s_{\alpha\beta}$ and
$\s^{CF}_{\alpha\beta}$ as the {\it disorder-averaged}
conductivities.\cite{lkgk}

Putting Eqs.(\ref{8989}) and (\ref{sxy}) together we obtain \eq
\frac{e^2}{2h}=\frac{\rxy}{\rxx^2+\rxy^2}=\frac{\rxy^{CF}+2\frac{h}{e^2}}
{(\rxx^{CF})^2+(\rxy^{CF}+2\frac{h}{e^2})^2}.\label{tr} \ee
After some trivial arithmetic we obtain \eq
\sxy^{CF}=\frac{\rxy^{CF}}{(\rxx^{CF})^2+(\rxy^{CF})^2}=-\frac{e^2}{2h},
\label{tr1} \ee so long as \eq
(\rxx^{CF})^2+(\rxy^{CF})^2 > 0. \label{condi} \ee The problem
lies in the fact that \Eq{scf} and \Eq{tr1} do not
agree.

Thus it seems that the notion of a Fermi liquid of composite
 fermions is incompatible with the particle-hole
symmetry. At this juncture the readers might wonder why
should we put so much weight on particle-hole symmetry.
After all there is no reason that such symmetry must exist in
real systems. To answer this question we quote
a recent experimental result of  Wong, Jiang and
Schaff.\cite{hw} In Ref.\cite{hw}  Wong {\it et al} measured $\rho_{xx}$
and $\rho_{xy}$ near $\nu=1/2$ in gated $GaAs/AlGaAs$
hetrostructures. What they found is that for a range of
carrier densities both $\rho_{xx}$ and $\rho_{xy}$ are
{\it temperature dependent} at $\nu=1/2$. However their
temperature dependence is such that $\s_{xy}=\rxy/(\rxx^2+\rxy^2)$ is
{\it temperature independent} and equals to
$e^2/2h$. Even leaving aside the issue of what is causing
$\s_{xy}=e^2/2h$, the very fact that $\s_{xy}=e^2/2h$
and $\rxx\ne 0$ is sufficient to give $\s^{CF}_{xy}=-e^2/2h$.
Recently Jiang has performed particle-hole
transformation (\Eq{ph}) on the data reported in Ref.\cite{hw}.
The result is entirely consistent with the
presence of particle-hole symmetry.\cite{hw1}

Since the derivation of $\s^{CF}_{xy}=-e^2/2h$
requires $(\rxx^{CF})^2+(\rxy^{CF})^2 >0$, and in the absence of
disorder $(\rxx^{CF})^2+(\rxy^{CF})^2$ could
vanish, it has been suggested that perhaps $\s^{CF}_{xy}=0$ in the
{\it zero disorder limit}. A difficulty with
this scenario is that once accepting $\s^{CF}_{xy}=0$ for no
disorder, one is forced to conclude that the
composite fermion Hall conductance jumps from $0$ to
$-\frac{e^2}{2h}$ upon the introduction of
{\it infinitesimal} amount of particle-hole symmetric disorder. Even
more bothersome is the fact that such jump
must persist at non-zero temperatures.\cite{note}

In any case our goal is to understand real systems for
which Wong {\it et al}'s result suggests
$\s^{CF}_{xy}=-\frac{e^2}{2h}$. In the following we
shall argue that $\s^{CF}_{xy}=-e^2/2h$ implies that the
(polarization) charge current carried by neutral
fermions does not have off-diagonal correlation. Since the last
statement is required by the fact that neutral
fermions are globally neutral, we believe that \Eq{tr1} holds even
in the absence of disorder.
\\

\noindent{\bf{III. The neutral fermion (dipole) theory}}
\\
\rm

The physical idea behind the neutral fermion
theory is as follow. Let us first consider a different problem where
we have a group of {\it distinguishable}
particles with identical mass at $\nu=1/2$.\cite{note1}
 If these
particles interact via a {\it sufficiently short-range
repulsive force}, the ground state wavefunction will be \eq
\Psi_{1/2}(z_1,...,z_N) =\prod_{(ij)}(z_i-z_j)^{2} \exp\{-\sum_k|z_k|^2/4\}. \label{wf} \ee Since \Eq{wf} is the
ground state in the absence of symmetry constraint,
it is the lowest-energy solution.

Of course, the symmetric wavefunction in
Eq.(\ref{wf}) is {\it not} allowed for electrons. Consequently electron's
Fermi statistics frustrates their energy
 minimization.\cite{priv} To quantify this frustration we view each
electron as a {\it boson} carrying one quantum of fictitious magnetic flux (Fig.2).

\begin{figure}[h]
\epsfysize=2.5cm\centerline{\epsfbox{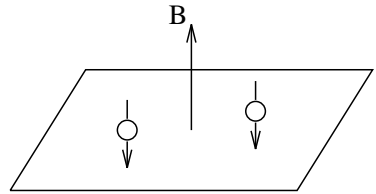}}
\vspace{20pt}
\caption{Each electron is viewed as a boson
carrying
 a quantum of fictitious magnetic flux. } \label{compboso}
\end{figure}

Were it not for the fictitious flux, the bosons would
have condensed into the Laughlin liquid described by
Eq.(\ref{wf}). Each fictitious flux quantum induces a
quasiparticle of charge $-1/2$ and statistics $\pi/2$. Due
to global charge neutrality a quasihole of opposite
charge are nucleated elsewhere. It turns out that the
quasiholes also have statistics $\pi/2$.\cite{note2}

Thus $\nu=1/2$ can be thought of as a liquid of
$\pm 1/2$ charged anyons floating on top of a Bose quantum Hall
liquid.\cite{lee} In Refs.\cite{lee} it is shown
that the action for this defect-liquid is given by \eq
S=S_{int}[-\gr\cdot\v P]+i\int d^2xdt[2\pi\v P\times \dot{\v P}-4\pi\v P\times\v j].
\label{act}
 \ee
 In the above
$\v P$ is the polarization density caused by the
 defects, and \eq \v j(\v x,t)=\sum_i\dot{\v r}_i(t)\delta(\v x-\v
r_i(t)), \ee where $\{\v r_i(t)\}$  are the
coordinates of electrons. For reason to be discussed shortly we
identify $\v j$ as the {\it composite fermion} current.\cite{erratum}
The partition function is given by \eq
Z=\int D[\{\v r_j\}]\int' D[\v P]e ^{-S}, \ee where $\int D[\{\v r_j\}]$
denotes the fermion-Feynman path integral
over $\{\v r_j(t)\}$, and $\int' D[\v P]$ denotes the function integral
 over $\v P(\v x,t)$ under the constraint
\eq -\gr\cdot\v P (\v x,t)=\sum_i\delta(\v x-\v r_i(t))-\rb. \label{cons} \ee
In this theory the total electric
current is the sum of the Hall current carried by the Bose quantum Hall
liquid, and the polarization current of
the defects. As the result\cite{lee} \eqa \s_{\al\beta}
=\frac{e^2}{2h}\epsilon_{\al\beta}+\s^N_{\al\beta}.
\label{nadd}\eea In \Eq{nadd} $\s^N_{\al\beta}$ is the
conductivity due to the polarization current \eq
j^N_{\mu}=(-\gr \cdot\v P, \dot{\v P}). \label{pc} \ee

If we {\it assume} that the Coulomb interaction binds
opposite-charged defects together, we arrive at a system of
dipoles. The polarization due to such dipoles is \eq \v P(\v x,t)
=\sum_i\v p_i(t)\delta(\v x-\v r_i(t)),
\label{dip} \ee where $\v p_i$ is the moment of the ith dipole.
By substituting \Eq{dip} into \Eq{act} and
\Eq{cons}, we can derive the action and the constraint reported
in Ref.\cite{shankar} and Ref.\cite{haldane}. It
is important to note that unlike ordinary dipoles, these dipoles obey Fermi
statistics.\cite{read0,shankar,lee,haldane,read}

Particle-hole symmetry also imposes a constraint
on the dynamics of dipoles. Combining \Eq{sxy} and \Eq{nadd} we
obtain \eq \s^N_{xy}=0. \label{phs} \ee \Eq{phs} is
easy to understand -- a liquid made up of (fermionic) dipoles
can not see the external magnetic field, hence has no
Hall conductance at $\v q=0$.
\\

\noindent{\bf{IV. The relation between the neutral
fermion and the fermion Chern-Simons theories}}
\\
\rm

There has been considerable discussions about
whether the neutral fermion and the fermion Chern-Simons theories
are actually the same.\cite{read,stern,stern1}
There are good basis for such suspicion. For
example the electron
density-density correlation function obtained
from neutral fermion theory is similar to that predicted by the
fermion Chern-Simons theory.\cite{lee,read,stern}
In addition, it has been shown recently that the neutral fermion
theory suffers similar infrared problems as the
fermion Chern-Simons theory.\cite{read,stern1} In the following we
shall prove that the neutral fermion theory is
in fact the fermion Chern-Simons theory formulated in the lowest
Landau level.

To see that we define \eq \v a\equiv -4\pi \zh\times\v P.
\label{def}\ee Substituting \Eq{def} into \Eq{act} and
\Eq{cons} we obtain
 \eq
S=S_{int}[\frac{1}{4\pi}\gr\times\v a] +i\int d^2xdt[\frac{1}{8\pi}
\v a\times\dot{\v a}+\v a\cdot\v j],
\label{actt} \ee and \eq \frac{1}{4\pi}\gr\times\v a (\v x,t)
=\sum_i\delta(\v x-\v r_i(t))-\rb. \label{conss} \ee
We recognize that Eqs.(\ref{actt},\ref{conss}) are the
first-quantized formulation of the fermion Chern-Simons
theory in the temporal gauge ($a_0=0$). We note that there
is no kinetic energy term $\int
dt\sum_i\frac{m}{2}|\dot{\v r}|^2$ because \Eq{actt} is
an action in the lowest Landau level.
\\

\noindent{\bf{V. The composite fermion Hall conductance
in the neutral fermion theory}}
\\
\rm

The equivalence of the neutral fermion and the fermion
 Chern-Simons theories does not say anything about the
validity of \Eq{scf}. To address that issue let us
concentrate on \Eq{actt} and \Eq{conss}. In the following we shall
demonstrate that due to\Eq{conss} there is no mixing
between the longitudinal and transverse
components of $\v a$.

Let us write \eq \v a=\v a_T+\gr\chi.\label{sep}\ee
By direct substitution it is simple to prove that in
the absence of boundary \eq S=S[\v a\ra\v a_T]-i\int d^2xdt\chi
[\gr\cdot\v j+\frac{1}{4\pi}
\gr\times\dot{\v a_T}].\ee
The last term vanishes by \Eq{conss} and the composite
fermion current continuity equation.

Such no-mixing condition should be respected if we integrate
 out the composite fermions first. To make the
connection to Ref.\cite{hlr} more transparent, let us restore
the gauge freedom in \Eq{actt}. The new action
read \eq S=S_{int}[\frac{1}{4\pi}\gr\times\v a] +i\int
d^2xdt[\frac{1}{8\pi}\epsilon_{\mu\nu\lambda}a_{\mu}\p_{\nu}a_{\lambda}
+ a_{\mu}j_{\mu}]. \label{actt1} \ee (We
note that after restoring the gauge freedom, the constraint (\Eq{conss})
is incorporated into the action.)

Now we are ready to integrate out the composite fermions. The
resulting action will depend on $a_{\mu}$ alone.
After gauge fixing, i.e. $a_0=0$, the result had better not contain a
longitudinal and transverse mixing term. In
order for that to be true a counter term must be generated by the composite fermions to
cancel second
term in \Eq{actt1}.  This requirement translated into
$\s_{xy}^{CF}=-\frac{1}{4\pi}$, or
equivalently, $\s_{xy}^{CF}=-\frac{e^2}{2h}$. The
 absence of the longitudinal and transverse mixing guarantees
that the polarization current of neutral fermions
fluctuates in a time-reversal-invariant fashion.
\\

\noindent{\bf{VI. The mixed behavior of composite fermions}}
\\
\rm

So far we have been focusing on the composite fermion
Hall conductance. It turns out that in order for the neutral
fermion theory to describe experiments, the composite-fermion
transverse current-current correlation $\Pi_{tt}(\v
q,\w)$ must be similar to that of a Fermi liquid.\cite{lee,read,stern,stern1}
The puzzle is why despite their
large Hall conductance, the composite fermions have
some aspect (i.e. $\Pi_{tt}(\v q,\w)$) of a Fermi liquid. In
the following we offer an explanation.

Let us define \eq j^{CM}_{\mu}\equiv j_{\mu}-j^N_{L,\mu}. \ee
In the above $j_{\mu}$ is composite fermion current,
and \eq j^N_{L,\mu}\equiv (-\gr\cdot\v P_L, \dot{\v P_L}), \ee
is the {\it longitudinal} component of the
polarization current. According to \Eq{cons} \eq j^{CM}_0=\rb. \ee
As the result $\v j^{CM}$ is a pure transverse
current. In this way the composite fermion current is decomposed into
a longitudinal and a transverse component:
\eq j_{\mu}=j^{CM}_{\mu}+j^{N}_{L,\mu}. \ee The reason that the
composite fermion current-current correlation
shows mixed behavior is due to the fact that while $<j^{CM}_{\mu}j^{CM}_{\nu}>$
 and $<j^{N}_{L,\mu}j^{N}_{L,\nu}>$
have time-reversal invariant correlation at $\v q=0$, $<j^{CM}_{\mu}j^{N}_{L,\nu}>$ does not.

In summary, the conduction at $\nu=1/2$ is through two
currents: a pure Hall current, and a
polarization current. The polarization current is
produced by a liquid of fermionic dipoles.
Composite fermions are constituents of these dipoles.
At long wavelength the dynamics of
neutral fermion is zero-field-like, while that of
composite fermions is not.
Consequently we believe that the composite fermions
do not form a
Fermi liquid.
\\
\vspace{0.1in}

Acknowledgement: DHL is supported in part by
DOE via Los Alomos National Laboratory and the Lawrence Berkeley
National Laboratory. He thanks Steve Kivelson
for valuable discussions.
\\
\vspace{0.1in}

\centerline{\bf BIBLIOGRAPHY}
\bibliographystyle{unsrt}

\begin{thebibliography}{99}
\bibitem{note1} Filling factor
is the ratio between the total number of
particle and total number of magnetic flux quanta. It does not require
the concept of occupation of Landau levels.
\bibitem{note2} The reason that the quasiparticle
statistics is $\pi/2$ instead
of $-\pi/2$ is due to  the contribution from
the statistical flux quantum.
\bibitem{erratum} In Ref.\cite{lee} I miss-identified the fermionic
current $j_{\mu}= (j_0,\vec{j})$ in Eq.(8). The correct identification is with the
composite-fermion (not the electron) current. This error causes two
necessary changes in the Ref.\cite{lee}: 1) the $\Pi_{lt}$ and
$\Pi_{tt}$ appear after Eq.(12) and in Eqs. (3)
and (14) should be identified with the
response functions of
composite-fermions, 2) The first
constraint in Eq.(12) requires $\sigma^{CS}_{xy}=-1/4\pi$, instead of
$\sigma_{xy}=1/4\pi$ as stated right after Eq.(12).
All other results of Ref.\cite{lee} remain unchanged.
\end{thebibliography}

 \end{multicols}
 \end{document}